\documentclass[11pt, a4paper,floatfix,fleqn]{revtex4}
\usepackage{vmargin,xcolor} 
\usepackage {version}
\usepackage {natbib}   \bibpunct{}{}{,}{s}{}{;}
\newcommand{\citea}[1]{\bibpunct{}{}{,}{s}{}{;}\cite{#1}}
\newcommand{\citeb}[1]{\bibpunct{[}{]}{,}{n}{}{;}\cite{#1}}

\newcounter   {comptapp} \renewcommand{\thecomptapp}{\Alph{comptapp}}
\usepackage[english]{babel} 
\usepackage[utf8]{inputenc} 
\usepackage{graphicx,float,setspace,amssymb,amsmath} 

\newcommand{\equa}[1]{\begin{eqnarray} \label{#1}} 
\newcommand{\auqe}{\end{eqnarray}} 
\newcommand{\tab}[1]{\begin{tabular}{#1}} 
\newcommand{\bat}{\end{tabular} \\ }

\newcommand{\gdsaut}{\vskip 1.5cm \noindent} 
\newcommand{\mt}[1]{\tilde{#1}}

\newcommand{\dd}{\hat{m}_i\hat{m}_j -3 (\hat{m}_i \hat{r}_{ij})(\hat{m}_j \hat{r}_{ij})}
\newcommand{\ku}{K^{(u)}}
\newcommand{\tria}[1]{\{\hat{x}_{\alpha #1}\}, \alpha = 1,3} 
\newcommand{\tri}[1]{\{\hat{x}_{\alpha #1}\}} 
 \newcommand{\be}{\beta} 
 \newcommand{\ep}{\epsilon} 
\newcommand{\de}{\delta} \newcommand{\gD}{\Delta}  
\newcommand{\s} {\sigma} \newcommand{\la}{\lambda}
 \newcommand{\tend}{\rightarrow} 
\providecommand{\abs}[1]{\left\vert#1\right\vert} 
 
\begin{document} 
\selectlanguage{english} 
\title 
{Magnetization of densely packed interacting magnetic nanoparticles with cubic and uniaxial anisotropies:
      a Monte Carlo study.
} 
\author 
{V.~Russier~$^{a)}$, C.~de-Montferrand~$^{b)}$, Y.~Lalatonne~$^{b)}$ and L.~Motte~$^{b)}$.} 
\affiliation
{~$^{a)}$~ICMPE, UMR 7182 CNRS and UPEC, 2-8 rue Henri Dunant 94320 Thiais, France.}
\affiliation
{~$^{b)}$~CSPBAT UMR 7244 CNRS and University Paris 13, 93017 Bobigny, France.}
\date {\today}
\thispagestyle{empty} 
\begin{abstract} 
The magnetization curves of densely packed single domain magnetic nanoparticles (MNP) are
investigated by Monte Carlo simulations in the framework of an effective one spin model.
The particles whose size polydispersity is taken into account are arranged in spherical clusters and 
both dipole dipole interactions (DDI) and magnetic anisotropy energy (MAE) are included in the total energy.
Having in mind the special case of spinel ferrites of intrinsic cubic symmetry, combined 
cubic and uniaxial magnetocrystalline anisotropies are considered with different configurations 
for the orientations of the cubic and uniaxial axes. It is found that the DDI, 
together with a marked reduction of the linear susceptibility are responsible for a damping 
of the peculiarities due to the MAE cubic component on the magnetization. As an application, 
we show that the simulated magnetization curves compare well to experimental results for 
$\gamma$--Fe$_2$O$_3$ MNP for small to moderate values of the field.
\end{abstract}
\maketitle 
%
\section {Introduction} 
\label{intro} 

Magnetic nanoparticles (MNP) assemblies present a fundamental interest in the development of nanoscale
magnetism research and are promising candidates in a wide range of potential applications going 
from high density recording to bio-medicine~\citea{skomski_2003, battle_2002, bedanta_2009, 
majetich_2006, bader_2006}.
Experimentally MNP can be obtained either as colloidal suspensions where the concentration can be varied
at will, embedded in non magnetic material where one can tune the interparticle interactions
or as powder samples where they form densely packed systems. 
The case of iron oxide particles, which are typical cubic spinel ferrites take a 
central place especially for bio-medical applications because of their biocompatibility and suitable superparamagnetic 
properties~\citea{pankhurst_2003,gupta_2005}.

The magnetic behavior of nanostructured materials or systems including nanoscale magnetic particles
is a multiscale problem since
the local magnetic structure within NP at the atomic site scale presents 
non trivial features~\citea{kodama_1996,labaye_2002,berger_2008,russier_2009} 
and the interactions between particles play an important role.  
A simplification occurs for NP of diameter below some critical value of typically few tens 
of nanometers since they then reach the single domain regime and can be described through an 
effective one spin model (EOS) where each NP is characterized by its moment and anisotropy energy.
However both the moment value and the anisotropy energy function are to be understood as effective 
quantities which take into account some of the atomic scale characteristics~\citea{morales_1999, 
tronc_2003, garanin_2003, kachkachi_2006}.
The EOS type of approach is a simplifying but necessary step for the description
of interacting MNP assemblies. In the framework of the EOS models, 
the total energy includes on the one hand the NP anisotropy energy through a one-body term and on the 
other hand the interparticle interactions. It is generally assumed for frozen systems of well 
separated NP that the leading term in the interparticle interactions is the dipolar interaction 
(DDI) between the macrospins which is totally determined once the NP saturation magnetization and 
the size distribution are known. Conversely modeling the anisotropy energy is not straightforward 
since in finite sized particles it comes from different origins.
The intrinsic contribution which stems form the bulk material is
{\it a priori} known experimentally without ambiguity. It
can be of either uniaxial or cubic symmetry according to the crystalline structure 
with anisotropy constants whose magnitude and even sign depend on temperature. 
In the widely studied case of oxide spinel ferrites at room temperature the intrinsic 
anisotropy is of cubic symmetry~\cite{coey_2010, birks_1950, bickford_1950, babkin_1984, smit_1959} 
with in general a negative constant $K_c$,
leading to the moment preferentially oriented along the $\{111\}$ directions of the crystallites. 
Then for NP not strictly spherical one has to add the shape anisotropy
term resulting from demagnetizing effect at the particle scale which for ellipsoidal NP is 
uniaxial with a shape anisotropy constant proportional to the NP volume~\cite{skomski_2003}.
Finally the finite size of the NP is the source of surface anisotropy 
resulting from symmetry breaking, surface defects or chemical bonding of the coating layer.
When modeled by a transverse anisotropy or the 
N\'eel surface anisotropy model~\citea{neel_1954}, the   
resulting non collinearities of the surface spins can be represented through a 
cubic term in the framework of the EOS~\citea{garanin_2003, kachkachi_2006}.
Concerning spherical iron oxide nanoparticles, the general experimental observation
is that the uniaxial anisotropy dominates  
with however a rather large dispersion in the effective annisotropy constant value 
$K_{eff}$~\citea{gazeau_1998,tronc_2003,roca_2006,dutta_2004,nadeem_2011,demortiere_2011,
tadic_2012}. 
Moreover a small value of $K_{eff}$ is interpreted as a small amount of crystalline defects within or at the 
nanoparticle surface~\citea{dutta_2004, tadic_2012,pereira_2010,pereira_2012}.
In any case the effective uniaxial anisotropy constant cannot be compared to the intrinsic, or bulk one,
since the latter corresponds to the cubic symmetry and is negative at room temperature.
Furthermore from the particle size dependence it is generally concluded that the uniaxial anisotropy 
is predominantly a surface anisotropy with a related constant $K_s=(d/6)K_{eff}$~\citea{gazeau_1998,
tronc_2003,demortiere_2011}. 

At the atomic scale the well known 
Néel model of pair anisotropy~\cite{neel_1954}
is often invoked to deduce surface anisotropy either in thin film geometry or in 3D NP.
In the framework of EOS approach, the deviation from the spherical shape translates in the Néel 
pair anisotropy model, in addition to the magnetic dipolar term responsible for the shape anisotropy, 
in a contribution with the same symmetry and proportional to the NP surface because of its short 
range character. This leads for ellipsoidal NP to a surface contribution of uniaxial symmetry. 
One has to keep in mind however that the Néel model although useful in the sense that it reproduces 
the correct description of the symmetry of the magnetic anisotropy, does not provide the physical 
understanding of the single ion anisotropy~\citea{skomski_2008}. Thus when dealing with spinel 
ferrites oxides as well as with Fe or Ni single domain nearly spherical nanoparticles in the 
framework of a EOS model, combined uniaxial and cubic anisotropies should be taken into account 
because of the intrinsic cubic anisotropy on the one hand, 
the shape (uniaxial) and surface contributions (uniaxial and/or cubic) on the other hand. 
 
In Ref.~\citeb{garcia-palacios_2000} the expansion of the linear and non linear susceptibility for
non interacting assembly with either uniaxial or cubic anisotropy has been performed with 
the result that when the 3 axes of the cubic contribution are randomly distributed both
the linear and the first non linear susceptibilities are anisotropy independent.
In Ref.~\citeb{usov_2012} the LLG equation is considered to calculate the hysteresis curve at 
vanishing temperature of non interacting NP, with randomly oriented cubic axes. The easy
axis of the uniaxial term is either fixed at conveniently chosen direction or randomly distributed.
In Ref.~\citeb{margaris_2012} an assembly of weakly interacting NP is considered both from
perturbation theory and MC simulations with cubic anisotropy relative to the same cubic axes 
for all the NP combined with an uniaxial anisotropy with a random distribution of easy axes. 

In the present work, we perform MC simulation of NP assemblies interacting through DDI 
with cubic and uniaxial contributions to the anisotropy energy. Having in mind the case of strongly 
interacting powder samples of NP dispersed at zero field, we consider the case of NP with cubic axes 
randomly distributed. The uniaxial easy axis on the other hand is either randomly distributed 
independently of the cubic axes or oriented along a particular crystallographic orientation of 
the particles for which two cases are considered, namely $\{100\}$ or $\{111\}$.
The main purpose of the present work is to investigate whether the cubic contribution to the anisotropy
leads to an observable deviation to the magnetization curve in the superparamagnetic regime.
We also revisit the consequences of the DDI in the strong coupling regime, in particular on the
linear susceptibility at low field, through the comparison of 
simulations performed either with free boundary conditions on spherical cluster or with 
periodic boundary conditions to simulate an infinite system.

In section~\ref{model}, we give the details of the model and explicit the different energy 
contributions. Section~\ref{results} is devoted to the results and the comparison with 
experimental results and we briefly conclude in section~\ref{conclusion}.
\gdsaut
\section{Model for densely packed assemblies } 
\label{model} 
We consider a EOS model with nanoparticles described as non overlapping spheres 
bearing at their center a permanent point dipole representing the uniform magnetization 
of the particle (macro spin). The moment of each particle is equal to its volume times 
the bulk magnetization, $M_s$, which means that no spin canting effect is explicitly taken 
into account. The particles are surrounded by a non magnetic layer of thickness $\gD$/2, 
representing the usual coating by organic surfactant molecules.
The particle diameters, $\{d_i\}$ are distributed according to a log-normal law 
defined by the median diameter $d_m$ and the standard deviation $\s$ of $ln(d)$,
\equa{lgn_1}
f(d) = \frac{1}{d \sqrt{2\pi}\s}\exp \left(-\frac{(\ln(d/d_m))^2}{2\s^2} \right) 
\auqe
In the following, we use $d_m$ 
as the unit of length, and the distribution function in reduced unit is totally determined
by the single parameter $\s$ which characterizes the system polydispersity.
When dealing with interacting particles, we mainly have in mind the case of lyophilized powders samples 
or high concentration nanoparticles assemblies embedded in non magnetic matrix.
Accordingly the coated particles are distributed in densely packed clusters whose external shape is spherical in order to
avoid the demagnetizing effects due to the system shape with the free boundary conditions. 
We emphasize that this NP configuration has an experimental justification
since upon drying the NP are likely to aggregate in spherical shaped large clusters
which has been confirmed from simulations \citea{lalatonne_2005}.
Moreover, we consider mainly the superparamagnetic regime, where we simulate only equilibrium 
magnetization curves corresponding to the static or infinite time measurements ($\tau_m\;\;\tend\;\infty$). 

We include only the leading terms of the anisotropy energy; the cubic symmetry contribution 
for particle say $i$ of moment $v(d_i)M_s \hat{m}_i$ can be written as 
\equa{an_cub1}
\frac{E_c^{(i)}}{v(d_i)} = K_c \left( m_{xi}^2m_{yi}^2 +  m_{yi}^2m_{zi}^2  + m_{xi}^2m_{zi}^2 \right)
                   = \frac{K_c}{2} \left( 1 - \sum_{\alpha = x,y,z} m_{\alpha i}^4  \right)
\auqe 
where we have used the unitarity of $\hat{m}_i$ in the second equality. 
Here and in the following hated letters denote unit vectors. In equation (\ref{an_cub1})
$m_{\alpha i}$ refer to the $\hat{m}_i$ components in the local cubic frame of the particle considered.
Let us denote by $\tria{i}$ this local cubic frame; dropping an irrelevant constant,
the total cubic anisotropy of the system can be written
\equa{an_cub2}
E_c = -\frac{K_c}{2} \sum_i v(d_i) \sum_{\alpha} (\hat{x}_{\alpha i} \hat{m}_i)^4
\auqe
The local axes $\tri{i}$ can be oriented in different ways according to the physical system under study; 
for non textured distributions of particles we have to consider a random distribution of the $\tri{i}$.
Most of our simulations are performed in the case. 
The effect of the texturation is nevertheless examined by considering that the 
$[111]$ directions $\{\hat{x}_{1}+\hat{x}_{2}+\hat{x}_{3}\}_i$ or the $\{\hat{x}_{3}\}_i$ axes are
confined in a cone along the $\hat{z}$-axis according to the following probability distribution for 
polar angles
\equa{an_cub3}
P(\theta) = C sin(\theta)exp(-(\theta/\s_{\theta})^2/2) ~;
\auqe
The configuration with the $\tri{i}$ fixed parallel to the system frame $(\hat{x}, \hat{y}, \hat{z})$ 
is also considered. This later case is the same as that~\citet{margaris_2012} considered 
for which a strong effect of the cubic term is obtained while~\citet{usov_2012} consider cubic axes 
randomly distributed. Although we have in mind particles of intrinsic cubic anisotropy, we are 
aware of a possible surface contribution to $E_c$ as shown in Ref.~\citeb{garanin_2003} 
resulting from the non collinearity of the surface spins; as a result the value of $K_c$ may 
differ form the bulk one.
The uniaxial term is proportional either to the volume $v(d)$ or to the surface $s(d)$ of the particle.
The volume part stems {\it a priori} from the shape anisotropy where for ellipsoidal particles 
$\ku_{sh}$, given by $J_s^2(1-3N_u)/(4\mu_0)$ with $J_s=\mu_0M_s$ and $N_u$, the demagnetizing factor
along the revolution axis, can be deduced from the knowledge of the aspect ratio $\xi$.
Notice that one can imagine easily a situation where the deviation from sphericity is not characterized 
by the same aspect ratio for all particles leading to a size dependence of $K_{sh}$. 
For instance one cannot rule out the situation where the deviation from sphericity follows from a major 
axis of the form $c = (d/2 + \delta)$ and minor axes $a$~=~$b$~=~$d/2$, with a size independent corrugation $\delta$. 
Then from the demagnetizing factor in the major axis direction 
\equa {N_c}
N_u = \frac{1 - \ep^2}{2\ep^3}\left[\ln\left(\frac{1+\ep}{1-\ep}\right) - 2\ep \right] 
    ~~ ; ~~ \textrm{with}~~\ep = (1-1/\xi^2)^{1/2} ~~ \textrm{and} ~~ \xi = c/a = 1 + 2\de/d ~, \nonumber
\auqe
the shape anisotropy may transform in a surface uniaxial anisotropy with an anisotropy constant given by 
$\ku_s \simeq J_s^2\delta/(30\mu_0)$ from an expansion of $\ku_{sh}$ at order $\delta/d$.

The easy axes $\{\hat{n}\}_i$ are either randomly distributed independently 
of the particles frame or aligned along one specified crystallographic axis of 
the crystallite. Different origin for such a easy axes distribution can be invoked. In the 
framework of the uniaxial anisotropy originating from the deviation to sphericity it corresponds 
to a preferential crystallographic orientation for crystallite growth, while in the framework of the 
uniaxial surface anisotropy this may result from a preferential crystallographic orientation for 
chemical bonding at the particle surface. We have considered two possibilities, namely 
$\hat{n}_i$~=~$\{001\}_i$ or $\hat{n}_i$~=~$\{111\}_i$.   

In the total energy, we include formally both surface and volume terms in the uniaxial contribution, 
with anisotropy constants $\ku_v$ and $\ku_s$ respectively and at most one of these is non 
zero in the simulations. The total energy thus includes the DDI, 
the one-body anisotropy term and the Zeeman term corresponding to the interaction with 
the external applied field $\vec{H}_a = H_a\hat{h}$.
Let $\{\vec{r}_i\}$, $\{v(i)\}$ , $\{\vec{m}_i\}$ and $\{\vec{n}_i\}$ denote the particles locations, 
volumes, moments and easy axes respectively. The total energy of the cluster reads 
\equa{ener_dip1}
E = \frac{\mu_0}{4\pi} \sum_{i < j} m_i m_j \frac{\hat{m}_i\hat{m}_j -3 (\hat{m}_i \hat{r}_{ij})(\hat{m}_j \hat{r}_{ij})}{r_{ij}^3}
    - \sum_i (\ku_v v(i) + \ku_s s(i))(\hat{n}_i\hat{m}_i)^2             \nonumber \\ \noindent
    - \frac{K_c}{2} \sum_i v(i) \sum_{\alpha} (\hat{m}_i\hat{x}_{\alpha i})^4
    - \mu_0 H_a \sum_{i} m_i \hat{m}_i \hat{h}
\auqe
$m_i$ are the moment magnitudes, 
$r_{ij} = \abs{\vec{r}_i - \vec{r}_j}$. It is worth mentioning that the consideration of the 
anisotropy term with a fixed easy axes distribution means that the magnetization relax according 
to a N\'eel process~\citea{du-tremolet_2000, coey_2010},
namely the particles are considered fixed while their moment relaxes relative to their easy axis.  
In the following we use reduced quantities; first the energy 
is written in $k_B T_0$ units, $T_0$ being a suitable temperature ($T_0 = 300K$ in the present work) 
and we introduce a reference diameter, $d_{ref}$. 
The reference diameter, $d_{ref}$ is a length unit independent of the size distribution, useful
for the energy couplings, and can be chosen from a convenient criterion independently of the actual structure
of the MNP assembly.
The reduced total energy is given by
\equa{ener_dip2}
\be_0 E &=& - \ep_{uv}^{(0)} \left( \frac{d_m}{d_{ref}} \right)^3 \sum_i d_i^{*3} (\hat{n}_i\hat{m}_i)^2
            - \ep_{us}^{(0)} \left( \frac{d_m}{d_{ref}} \right)^2 \sum_i d_i^{*2} (\hat{n}_i\hat{m}_i)^2 \nonumber \\
        &-&   \frac{\ep_{c}^{(0)}}{2} \left( \frac{d_m}{d_{ref}} \right)^3 \sum_i d_i^{*3} \sum_{\alpha} (\hat{m}_i\hat{x}_{\alpha i})^4 
            - \ep_d^{(0)} \left(\frac{d_m}{d_{ref}}\right)^3 
            \sum_{i < j} d_i^{*3}d_j^{*3} \frac{\dd}{r_{ij}^{*3}}     \nonumber \\
        &-& h \sum_{i} d_i^{*3} \hat{m}_i \hat{h} ~,                 
\auqe
with 
\equa{coupl_cste}
            \ep_{uv}^{(0)} = \be_0 K_v^{(u)} v(d_{ref}) ~;~ \ep_{us}^{(0)} = \be_0 K_s^{(u)} s(d_{ref}) ~;~ \ep_{c}^{(0)} = \be_0 K_c v(d_{ref})   \nonumber \\
            \ep_d^{(0)} = \frac{\be_0 \mu_0}{4\pi} (\pi/6)^2 M_s^2d_{ref}^3 ~;~
      h = \be_0\mu_0 M_s (\pi/6) d_{m}^3 H_a \equiv \left(\frac{d_m}{d_{ref}}\right)^3 \frac{H_a}{H_{ref}}
\auqe
where $\be_0 = (k_B T_0)^{-1}$ and the stared lengths are in $d_m$ unit.
The dimensionless dipolar coupling constant is then 
$\ep_d=(d_m/d_{ref})^3\ep_d^{(0)}$ 
and the dimensionless anisotropy constants are $\ep_{uv}=(d_m/d_{ref})^3\ep_{uv}^{(0)}$,
$\ep_{us}=(d_m/d_{ref})^2\ep_{us}^{(0)}$, 
and $\ep_{c}=(d_m/d_{ref})^3\ep_{c}^{(0)}$ for the volume and surface uniaxial and cubic contributions 
respectively. 
The reference diameter, $d_{ref}$ can be chosen such that 
$\ep_d(d_m=d_{ref})\equiv \ep_d^{(0)}$ = 1 ;
the reduced external field $h$ coincides
with the usual Langevin variable 
at temperature $T_0$ for a monodisperse distribution with $d$ = $d_m$
In equation (\ref{coupl_cste}), we also introduce
the reference external field, $H_{ref}$ for convenience. 

Concerning the structure in position, the nanoparticles surrounded by their coating layer of 
thickness $\gD$/2 form an 
assembly of hard spheres of effective diameters $\{d_i + \gD\}$
which are arranged in large densely packed clusters with either a random or 
a well ordered simple structure (simple cubic lattice)
The clusters are built as in Ref.~\citeb{russier_2012}.
First a large stacking of the coated spheres is made in a 
parallepipedic box with the desired structure, random or well ordered. 
Once this first step is performed, we cut within the global stacking the cluster we want to 
study by imposing both the external shape and the number of 
particles $N_p$, with typically $N_p \simeq$ 1000.
Because of the coating layer of thickness $\gD$/2  the closest distance of 
approach between particles $i$,~$j$ is shifted from $(d_i + d_j)/2$ to $(d_i + d_j)/2 + \gD$ and 
we therefore define an effective dipolar constant corresponding the particles uncoated at contact
 
\equa{ed_eff}
\ep_d^{eff} = \ep_d\frac{1}{(1 + \gD/d_m)^3} = \ep_d\frac{\phi}{\phi_m} \varphi(\s, \gD/d_m)
\auqe
where $\phi$ is the volume fraction and $\phi_m=\phi(\gD=0)$ is the maximum value of $\phi$
for a given structure. The function $\varphi$ (see appendix) in equation (\ref{ed_eff}) is equal to 1 for $\s=0$
and remains very close to 1 for $\s<0.1$.
Two systems differing by $\gD$ or $\s$ and characterized by the same value of $\ep_d^{eff}$ correspond to 
the same intensity of DDI. Notice that for weak polydispersity, $\ep_d^{eff}$ is related to the parameter $y$ widely 
used in the works dealing with the dipolar hard sphere fluid (DHS) which in our notations reads 
$y\;=\;8\beta^*\phi\ep_d/3$.

Although we do not limit our simulations to a specific experimental system,
we have in mind iron oxide NP to guide our choice of the physical parameters entering the model.
At room temperature the bulk anisotropy constant is $K_c$~=-11.0 to -13.0 kJm$^{-3}$ and 
-4.70 kJm$^{-3}$ for magnetite and maghemite respectively. The saturation magnetization for these 
two materials are quite close and lead to $J_s\;\sim$ 0.50T. 
Therefore in this work we use $J_s$~=~0.50\;T, which corresponds to 
$d_{ref}$~=~10~$nm$ when this later is fixed from $\ep_d(d_{ref})$~=~1.0. 
The shape anisotropy constant can be estimated for ellipsoidal NP once the aspect ratio is known; using
$J_s=0.5T$ we get $\ku_{sh}\simeq{50}(1-3N_l)$~kJm$^{-3}$ which leads to 
$\ku_{sh}~<$~7.0 kJm$^{-3}$ for NP characterized by an aspect ratio $\xi<1.20$, and accordingly $\ep_{uv}^{(0)}=0.95$.
Concerning the surface anisotropy constant, we consider the experimental values ranging from
$K_s$~=~5.5~10$^{-6}$~Jm$^{-2}$ to $K_s$~=~2.7~10$^{-5}$~Jm$^{-2}$
for maghemite~\citea{gazeau_1998,tronc_2003,tadic_2012}; thus $K_s$~=~2.7~10$^{-5}$~Jm$^{-2}$ is considered somewhat
as an upper bond for iron oxide NP. Notice that if we consider the deviation from sphericity resulting from 
a size independent corrugation $\delta$ as outlined above, we get $K_{sh}^{(s)}~\simeq$~2.70~10$^{-5}$~Jm$^{-2}$
with $\delta~$=2 $nm$, which corresponds to an aspect ratio ranging from 1.4 to 1.2 for NP of diameter ranging from 
10 to 20 $nm$.
The corresponding values of the reduced parameters for Iron oxide NP of {\it c.a.} 10 to 20 $nm$ in diameter are 
summarized in table \ref{tab_para}. In the following we can consider that a characteristic value for the uniaxial
anisotropy is about $\ep_{uv}\sim~5$ while a maximum value for the cubic anisotropy constant is 
$\abs{k_c}~=~15$.

\vskip 0.05\textheight

\begin{table}[h]
\vskip 0.02\textheight
\begin{tabular}{|c|c|c|c|c|c|}
\hline
  $d_m/d_{ref}$       &  1     & 1.20 & 1.33  &  1.71 &   2.0   \\ \hline
  $\ep_{c}~^{a)}$     & ~-1.65~&~-2.85~&~-3.88~&~-8.25~&~-13.20~\\    
  $\ep_{c}~^{b)}$     &  -0.60 & -1.05 & -1.41 & -3.0  &  -4.8  \\ \hline
  $\ep_{uv}~^{c)}$    &  0.625 &  1.08 &  1.47 &  3.13 &   5.00 \\ 
  $\ep_{uv}~^{d)}$    &  1.00  &  1.73 &  2.37 &  5.0  &   8.00 \\ \hline
  $\ep_{us}~^{e)}$    &  2.05  &  2.95 &  3.62 &  6.0  &   8.20 \\ \hline
  $\ep_d$             &  1.0   &  1.73 &  2.37 &  5.0  &   8.00 \\ \hline
  $\ep_d^{eff}~^{f)}$ &  0.60  &  1.09 &  1.56 &  3.6  &   6.00 \\ 
  $\ep_d^{eff}~^{g)}$ &        &       &       &       &   1.56 \\ \hline
\end{tabular}
\caption{\label{tab_para}
Reduced values for the parameters of the model corresponding to iron oxide NP.
with  $K_c$ = - 13 kJm$^{-3}$ $^{a)}$ ; -4.7 13 kJm$^{-3}$ $^{b)}$; 
an aspect ratio $\xi$ = 1.135 $^{c)}$; or 1.20 $^{d)}$;
or surface anisotropy constant $K_s$ = 2.70 10$^{-5}$~Jm$^{-2}$ $^{e)}$.
$\ep_d^{eff}$ from equ. (\ref{ed_eff}) with $\gD$~=2~nm $^{f)}$ or 14.5~nm $^{g)}$.
}
\end{table} 
\vskip 0.05\textheight

The main effect of the DDI on the magnetization curve at low field 
is a strong reduction of the
initial slope of $M(H)$ versus $H$, namely of the linear 
susceptibility~\citea{chantrell_2000, russier_2012}. This is directly related to
the well known plateau in the FC magnetization in terms of the temperature
for $T\;<\;T_B$ occurring in strongly interacting NP and also the the plateau in the
$\chi(d_m)$ curve obtained in Ref.~\citeb{russier_2012} for densely packed clusters of NP.
Indeed, in the absence of anisotropy, the hamiltonian (\ref{ener_dip2}) can be
rewritten, with $h_r~=~H_a/H_{ref}$
\equa{ener_dip_3}
\beta E = \beta^*\beta_0 E &=&  
       - \beta^* \ep_d^{(0)} \left(\frac{d_m}{d_{ref}}\right)^3 \frac{1}{(1 + \gD/d_m)^3}
            \sum_{i < j} d_i^{*3}d_j^{*3} \frac{\dd}{(r_{ij}^*/(1 + \gD/d_m))^{3}}                 \nonumber \\
       &-& h_r \beta^* \left(\frac{d_m}{d_{ref}}\right)^3 \sum_{i} d_i^{*3} \hat{m}_i \hat{h}  
\auqe
where we have introduced the geometrical sum of the reduced DDI of the most concentrated cluster ($\gD$~=0)
of the structure considered, namely where particles can get at contact which thus depends neither on $d_m$
and $\gD$. 
Introducing the dimensionless variable $\la~=~\beta^*(d_m/d_{ref})^3$ we get
\equa{ener_dip_4}
\beta E = 
       - \la \ep_d^{eff (0)} \sum_{i < j} d_i^{*3}d_j^{*3} \frac{\dd}{(r_{ij}^*/(1 + \gD/d_m))^{3}} 
       - h_r \la \sum_{i} d_i^{*3} \hat{m}_i \hat{h}
\auqe
From equation~(\ref{ener_dip_4}), we can conclude that, when the field is vanishingly small, 
the leading contribution to the
magnetization linear in $h_r$ depends on $h_r$ and $\beta^*$ only through $\la h_r$ and $\la \ep_d^{eff (0)}$.
More precisely, $M/M_s~\simeq~\la h_r f(\la \ep_d^{eff (0)})$ with $f$ a scaling function. 
Thus the linear susceptibility, $\chi~=~\partial M/\partial H_a$~$\simeq~M/H_a$ at vanishing $H_a$
must be in the form
\equa{chi_1}
\chi = \frac{M_s}{H_{ref}} \la f(\la \ep_d^{eff (0)}) 
\auqe
In the limit of zero coupling $f(x=0)$ is a finite constant and we recover the Langevin result,
$\chi\propto\la$ = $(6/\pi)(T_0/T)v(d_m)$. In the interacting system, the strong coupling 
limit $\la\ep_d^{eff (0)}>>1$ or equivalently $y>>1$ is obtained through the increase of either 
the DDI coupling, $\ep_d^{(0)}$ or $\beta^*$ (decrease of $T$). In this case, the limiting value 
of the susceptibility can be obtained. We note that the linear susceptibility we deal with is 
the external one, relating the magnetization to the external, or applied field $H_a$ and since we 
consider the magnetization per unit magnetic volume, the magnetization per unit volume is 
$M_v$~=~$M\phi$. Thus the internal field is related to the external one through 
$H_i=H_a-D_h{\phi}M$ where $D_h$ is the demagnetizing factor of the sample in the direction of 
the field. Hence we can relate $\chi$ to the internal susceptibility, $\chi_i$ through the usual 
way~\citea{coey_2010}
\equa{chi_i}
\chi = \frac{\chi_i}{1 + D_h\phi \chi_i}
\auqe
We can also introduce the relative permeability, $\mu=(1+\phi\chi_i)$ to get
\equa{chi_e}
\phi\chi = \frac{\mu - 1}{1 + D_h(\mu - 1)}
\auqe
In the case of a spherical system as those considered here, $D_h\;=\;1/3$ and equation (\ref{chi_e}) reads
\equa{chi_eb}
\phi\chi = \frac{3(\mu - 1)}{\mu + 2}
\auqe
%
It is worth mentioning that $\chi$ is related to the moment fluctuations through the fluctuation-dissipation 
theorem~\citea{allen_1987}  as already used in~\citeb{russier_2012}. We have in an isotropic system 
\equa{fluctu1}
  \frac{\partial(M/M_s)}{\partial h} = \chi_r 
  = \beta^* \frac{N \bar{v}}{3 v(d_m)} 
    \left( 
         \frac{\left< \left( \abs{ \Sigma \vec{m}_i } \right)^2 \right> }{ \left( \Sigma m_i \right) ^2 } 
       - \frac{ \abs{ \left<  \Sigma \vec{m}_i \right> }^2 }{ \left( \Sigma m_i \right) ^2 } 
    \right)
    \equiv \beta^* \frac{\bar{v}\;g}{3 v(d_m)} 
\auqe
which introduces the factor $g$ and where $\bar{v}$ is the average value of the particle volume over the
distribution function. From equation (\ref{fluctu1}) we rewrite (\ref{chi_eb}) in the equivalent form
\equa{chi_ec}
\frac{3(\mu - 1)}{\mu + 2} = \phi\chi = 8\phi\ep_d\beta^* \frac{\bar{v}}{v(d_m)} g
\auqe
Now in the strong coupling limit we expect the system to reach a ferromagnetic transition as is the case for
the DHS fluid~\citea{wei_1992,weis_1993}. In this limit the permeability $\mu\tend\infty$  and a limiting value for $\chi$ 
and thus a plateau in the FC magnetization when the temperature is decreased is obtained with, from
equation (\ref{chi_eb})
\equa{chi_lim}
    \chi \rightarrow \frac{3}{\phi} 
                                    ~~ \text{ or   } ~~ \tilde{\chi} \rightarrow  \frac{3}{8\ep_d^{(0)}\phi}
                                    ~~ \text{ with } ~~ \mt{\chi} = \frac{H_{ref}}{M_s} \chi
\auqe
This is quite well reproduced by the present simulations (see section~(\ref{results})) 
and in total agreement with the behavior of $\mt{\chi}$ in
terms of the particle size $d_m$ we obtained in Ref.~\citeb{russier_2012} in the quasi 
monodisperse case where $\varphi\simeq1$ which is easily deduced from (\ref{chi_lim}) 
by writing $\phi$ in terms of $\gD/d_m$
\equa{chi_lim2}
\mt{\chi} \tend \frac{\varphi~(1 + \gD/d_m )^3}{8\ep_d^{(0)}\phi_m}
\auqe
It is important to note that equation (\ref{chi_eb}) is the well known relation between the 
dielectric constant and the polarization susceptibility in the DHS fluid in the case of an 
infinite spherical system embedded in vacuum, {\it i.e.} surrounded by a medium 
of dielectric constant $\ep_s=1$. Indeed the magnetic permeability plays the role of the 
dielectric constant of the DHS and the polarization susceptibility is related to the 
fluctuations or the Kirkwood factor $g_K(\ep_s)$, equivalent to the factor $g$ introduced 
above; in the monodisperse case, with the dielectric constant, $\ep$, in place of $\mu$
the DHS satisfies~\citea{allen_1987,frenkel_2002}
\equa{diel_1}
 \frac{\mu - 1}{\mu + 2} = yg_K(\ep_s = 1)   ~~ ; \text{ or } ~~ \mu - 1 = 3yg_K(\ep_s = \infty)  
\auqe
Notice that the second equation (\ref{diel_1}) is the equivalent of (\ref{chi_e}) written for $D_h=0$
and corresponds to the case where either through the boundary conditions ($\ep_s=\infty$) or the 
system shape ($D_h=0$) the system can be uniformly polarized.
Equation (\ref{diel_1}) is strictly equivalent to (\ref{chi_ec}) since in the present model we have, in 
the monodisperse case, $\chi_r=\beta^*g/3$. The DHS undergoes a ferromagnetic transition at which 
the dielectric constant diverges and as a result~\citea{weis_1993,weis_2005,weis_2006}, one expects a limiting value for the Kirkwood factor 
$g_K(\ep_s~=~1)\tend~1/y$ and accordingly $\chi_r\tend\beta^*/(3y)$ 
or $\chi\tend3/\phi$ in agreement with equation (\ref{chi_lim}).

The plateau in the FC magnetization at low temperature and low field is a behavior observed
in the framework of the FC/ZFC procedure~\citea{vargas_2005,caruntu_2007,tan_2008,nadeem_2011,pereira_2012} 
generally related to a collective behavior of the dipoles leading to a frozen state.
Here, by analogy with the known behavior of the DHS fluid, we relate this plateau to the approach 
of the onset of the ferromagnetic transition
at least for $\s$~$<<$~1 and in the absence of MAE.
We emphasize that as can be deduced from equation~(\ref{chi_eb}), in the case of a spherical system 
surrounded by vacuum, $\chi$ becomes nearly independent of $\mu$ when $\mu$ increases beyond a 
sufficiently high ($\mu~\sim~35$) but still finite value. As a result $\chi$ gets close to its 
limiting value before the ferromagnetic transition.

The Monte Carlo simulations are performed according to the usual Metropolis 
scheme~\citea{binder_1997,allen_1987,frenkel_2002}. The trial move
of each moment is performed within a solid angle centered on its old position. Since we seek equilibrium
configurations, the maximum solid angle of the move is only restricted by the acceptance ratio, 
$R$~$\sim$~0.35--0.50. Moreover we use a annealing scheme at all values of the field in the range
where we expect an hysteresis. The averages are performed on 10 to 30 independent runs (up to 70 runs 
for low temperature and/or large DDi couplings) with 3\,10$^4$ to 4\,10$^4$ thermalisation MC
steps followed by another set of 3\,10$^4$ to 4\,10$^4$ MC steps to compute the averages. 

\section {Results}
\label {results}
\subsection* {Non interacting system}
\begin{figure}   
\includegraphics[width = 0.75\textwidth , angle = -00]{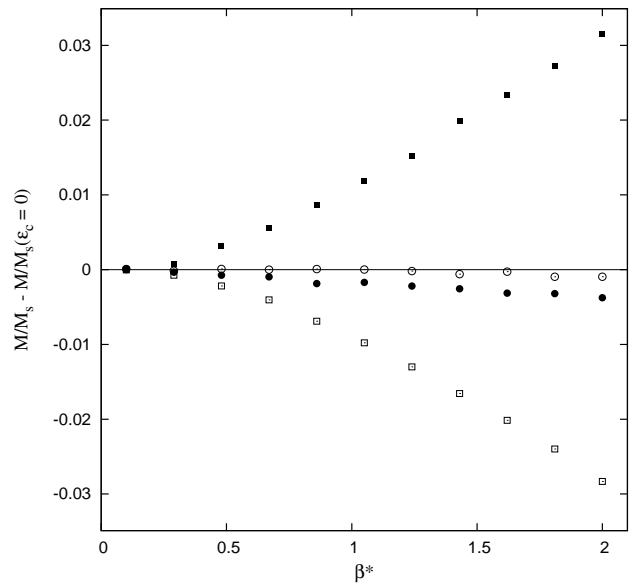}
\caption{\label{ed_0_ek_0_rand_fix_low_h}
Deviation of the reduced magnetization $M/M_s$ due to MAE at $h$ = 0.20 for a non interacting system with 
cubic anisotropy. Polydispersity: $\s$~=~0.28.
Cubic axes randomly distributed and $\ep_c$~=15, solid circles; $\ep_c$~=~-15, open circles. Cubic axes
fixed and parallel to the system frame with $\ep_c$~=~15, solid squares; $\ep_c$~=~-15, open squares.
} 
\end{figure}

In this section we deal with the case free of DDI. 
We first have checked that as $h\,\tend\,0$ 
with volume uniaxial MAE and a random easy axes distribution  the 
linear susceptibility is $\ep_{uv}$ independent while with cubic MAE and
randomly distributed axes, both the linear and the first
non linear susceptibilities are $k_c$ independent and accordingly we get a 
nearly $k_c$ independent M(h) beyond the very vicinity of $h$\,=\,0. This is shown in 
figure~\ref{ed_0_ek_0_rand_fix_low_h} in terms of the inverse reduced temperature $\be^*$. 
Moreover we also check in 
figure~\ref{ed_0_ek_0_rand_fix_low_h} that the deviation of $M(h)$ relative to the isotropic 
case is negative whatever the sign of $k_c$ with the random distribution of cubic axes. 
This is no more the case when the cubic axes of the particles are fixed where on the one 
hand only the linear susceptibility is $k_c$ independent and on the other hand the sign of 
$(M(h,k_c)-M(h,k_c=0))$ depends on the sign of $k_c$. The same result holds when $\ep_d\neq\,0$.

\begin{figure}   
\includegraphics[width = 0.75\textwidth , angle = -00]{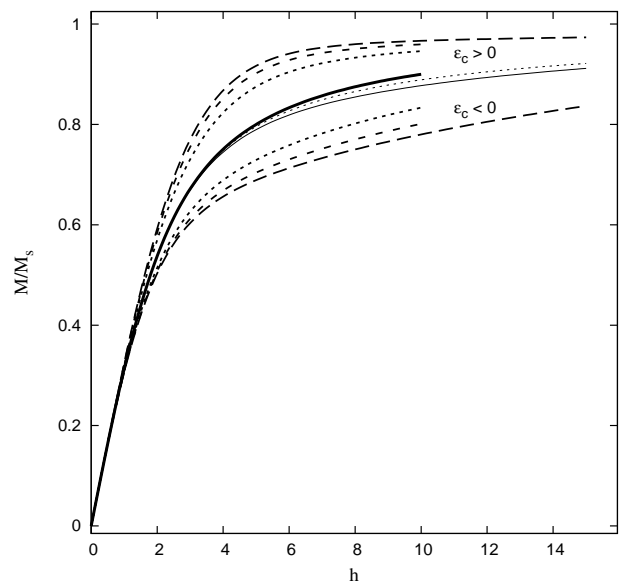}
\caption{\label{mh_cs_ed_0_ek_0_fix_rand_kcpm}
Magnetization curve for a monodisperse non interacting system with cubic anisotropy. The cubic anisotropy
axes are fixed along the system frame with $\ep_c$~=~$\pm$ 15 long dashed; $\pm$ 12 dashed; and $\pm$ 8 
short dasched. The sign of $\ep_c$ is as indicated. The case with random distribution of the cubic axes
is shown for comparison with $\ep_c$~=~15, thin solid line; and $\ep_c$~=-15, thin dotted line. The thick solid line
is the reference $\ep_c$~=~0 case. $\be^*$~=~1.   
}
\end{figure}

For randomly distributed cubic axes, the cubic MAE has only a negligible effect on the $M(h)$ curve. 
On the opposite, as shown in figure~\ref{mh_cs_ed_0_ek_0_fix_rand_kcpm}, when the cubic 
axes are fixed along the system frame, the cubic MAE has a strong effect on the $M(h)$ curve. 
Moreover, as noted above in the low field region, the sign of the anisotropy induced 
deviation of $M(h)$ depends on the sign of $k_c$. 
This is expected since a positive value 
of $k_c$ will favor the principal frame directions for the moments; for an applied field 
along one of these directions, say $\hat{h}$\,=\,$\hat{z}$, $k_c>0$ leads to a positive 
deviation of $M(h)$ and {\it vice versa}. The results displayed in 
figure~\ref{mh_cs_ed_0_ek_0_fix_rand_kcpm} are in agreement with those of 
Ref.~\citeb{margaris_2012} (notice that our $k_c$ corresponds to $w/2$ of
Ref.~\citeb{margaris_2012}). 

\begin{figure}[h]   
\includegraphics[width = 0.75\textwidth , angle = -00]{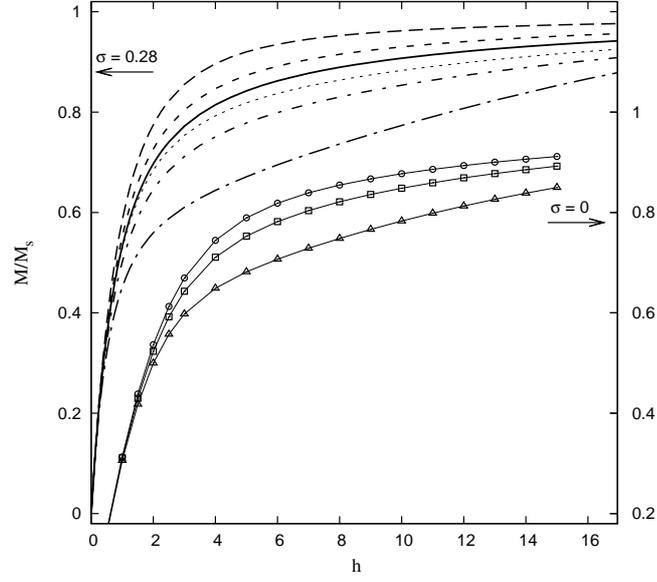}
\caption{\label{text_111_z_0.28_ed0_ek0_kcpm15}
Magnetization curve for non interacting system with cubic anisotropy, $\abs{\ep_c}$~=~15 and $\be^*$~=~1.
The $[111]$ direction of the cristallites are prefentially oriented along the $z$ axis (which is also the
direction of the field) with the probability distribution of equation~(\ref{an_cub3}). Polydisperse case 
($\s$~=~0.28) with $\ep_c$~=~-15 and $\s_{\theta}$~=~0.015, long dash; $\pi/10$, short dash; $\pi/2$, solid line.
Same with $\ep_c$~=~15 and $\s_{\theta}$~=~$\pi/2$, dotted line; $\pi/10$, short dash dot; 0.015, long dash dot.
Monodisperse case ($\s$~=~0) with $\ep_c$~=~15 and 
$\s_{\theta}$~=~0.015, open triangles; $\pi/10$, open squares; $\pi/2$, open circles.
}
\end{figure}

The effect of the texturation through the preferential orientation along the $\hat{z}$-axis 
of the crystallites $[111]$ direction according to the probability density (\ref{an_cub3}) 
is shown in figure~\ref{text_111_z_0.28_ed0_ek0_kcpm15} for the polydisperse and monodisperse 
cases.

\begin{figure}[h]   
\includegraphics[width = 0.75\textwidth , angle = -00]{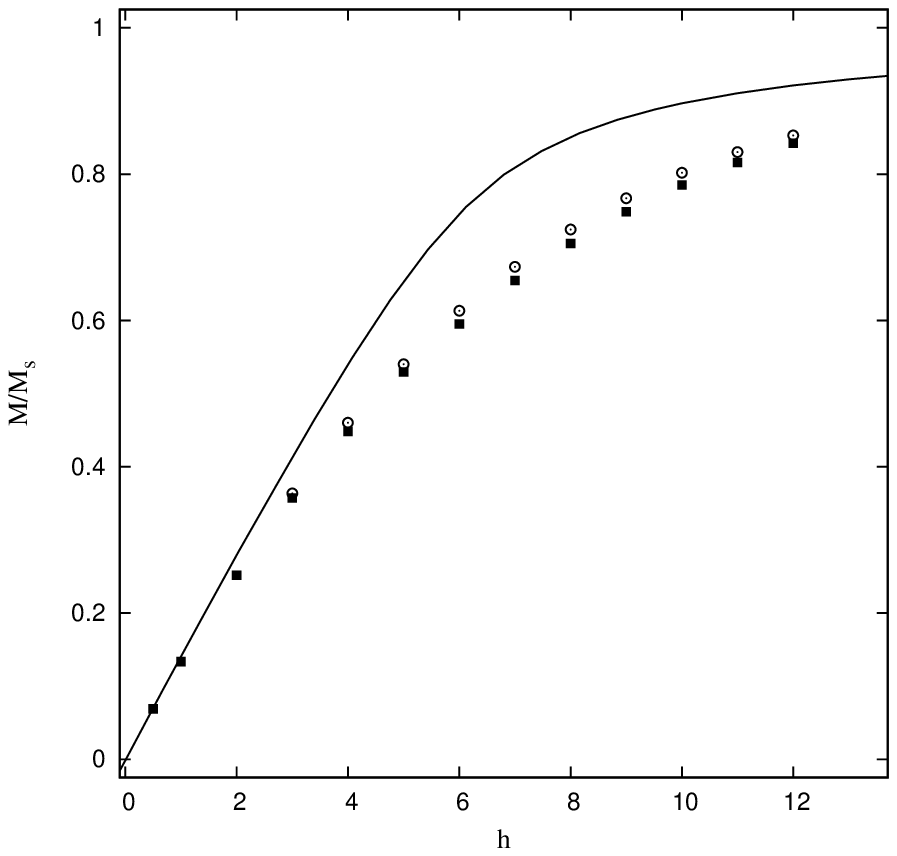}
\caption{\label{surf_vol_unax}
M(h) for an interacting system characterized by $\ep_d$~=~2.37, $\gD/d_{ref}$~=~0.20, $d_m/d{ref}$~=~1.33 and $\be^*$~=~1.
Without anisotropy: solid line. In the presence of uniaxial anisotropy with $\ep_{uv}$~=~5.64 and $\ep_{us}=0.0$, 
solid squares; $\ep_{uv}$~=~0.0 and $\ep_{us}$~=~6.88, open circles.
(The value $\ep_{us}$~=~6.88 corresponds to~$\ep_{uv}(d_3^*(\s)/d_2^*(\s))$ with $\ep_{uv}$~=~5.64, $d_n^*$ is the $n$-th 
moment of the diameter distribution function.)
}
\end{figure}

Concerning the uniaxial anisotropy, we note that the surface contribution can be 
very well approached by the volume term with the introduction of an effective volume 
uniaxial constant, $\ep_{uv}^{eff}$ taking into account the polydispersity. In 
equation~(\ref{ener_dip2}), we rewrite the uniaxial energy terms by introducing the 
reduced $n-th$~order moments $d_n^*$ of the diameter distribution function and under the hypothesis that
$(\sum\,d_i^{*n}(\hat{n}_i\hat{m}_i)^2)/d_n^*$ is independent of $n$ at least for $n\leq3$
we get
\equa{kv_eff}
\ep_{uv}^{eff} = \frac{d_2^*}{d_3^*} \ep_{us}
               = \exp(-5\s^2/2) \ep_{us}
\auqe
where we have used the analytical result for the $d_n^*$ of the lognormal law. The same 
conclusion holds in presence of DDI; in figure~\ref{surf_vol_unax} we compare the deviation 
of $M(h)$ due to the surface uniaxial MAE with that due to the volume uniaxial MAE with 
$\ep_{uv}~=~\ep_{uv}^{eff}$ taken from~(\ref{kv_eff}) in the case of a polydisperse 
interacting system.

\begin{figure}[h]   
\includegraphics[width = 0.75\textwidth , angle = -00]{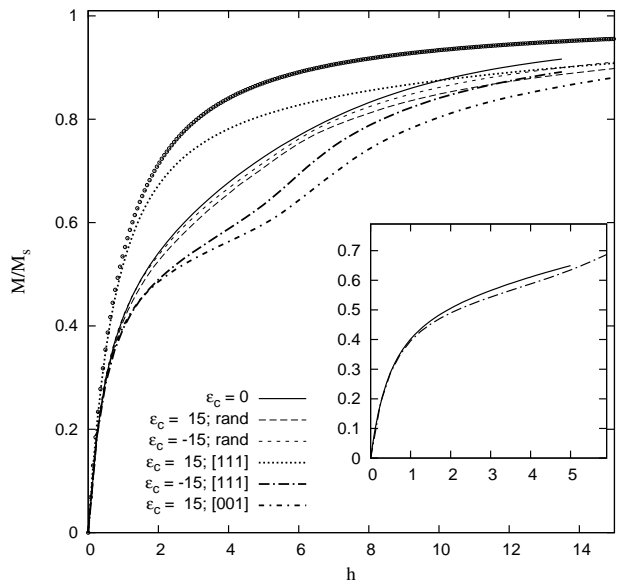}
\caption{\label{comb_ek_kc_noddi}
Magnetization curve for non interacting system with uniaxial and cubic anisotropies with $\be^*$~=~1, $\ep_{uv}=5$ ,
$\ep_{us}=0$ and $\abs{\ep_c}=15$. Polydispersity : $\s=0.28$. Open circles: case free of anisotropy for comparison.
$\ep_c$ and easy axes distributions as indicated.
Inset : comparison of the $M(h)$ curves for $\ep_{uv}=5$ and $\ep_c=-15$, long dash dotted line and for $\ep_{uv}=8$ and 
$\ep_c=0$, solid line.
}
\end{figure}

We now consider the case of combined uniaxial and cubic anisotropies. The result is shown
for a typical set of parameters, $\ep_{uv}=5$ and $\abs{k_c}=15$ in 
figure~\ref{comb_ek_kc_noddi}. As is the case when only the cubic anisotropy 
is taken into account, we find that the effect on $M(h)$ of the cubic anisotropy with 
random distributed cubic axes is very small when the uniaxial easy axes are also randomly
distributed and uncorrelated from the cubic ones. 
This is no more the case when, still for a random distribution of cubic axes, the easy axes
$\{\hat{n}\}_i$  are along a specified crystallographic orientation of the crystallites. 
The cubic MAE enhances the uniaxial one when $\ep_c > 0$ and
$\{\hat{n}\}_i$ = $[001]$, or when $\ep_c < 0$ and $\{\hat{n}\}_i$\,=\,$[111]$.
This is qualitatively expected since then the two components of the MAE 
tend to favor the same local orientation for the moment.

\begin{figure}[h]   
\includegraphics[width = 0.75\textwidth , angle = -00]{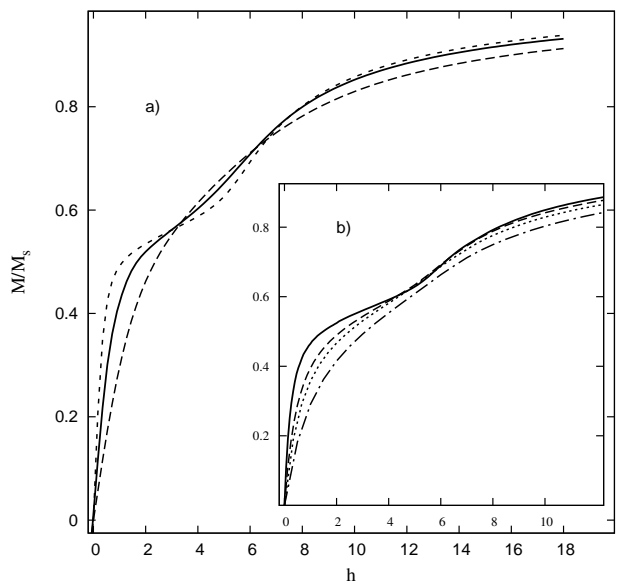}
\caption{\label{shoulder} 
Reduced magnetization for a non interacting system with $\ep_{uv}$~=~5.0, $\ep_c$~=~-15,
cubic axes randomly distributed, easy axes along the $[111]$ NP cristallographic orientations and different values
of the reduced inverse temperature $\beta^*$. 
$\beta^*$~=~0.5, dash dotted line; 0.75, dotted line; 1.0, long dashed line; 2.0, solid line; 4.0 short dashed line.
{\it a)} monodisperse system ($\s$~=~0); {\it b)} polydispersity $\s$~=~0.28.
}
\end{figure}

A shoulder in $M(h)$ is clearly observed when $\{\hat{n}\}_i=[001]$ and $\ep_c>0$ or
$\{\hat{n}\}_i$\,=\,$[111]$ and $\ep_c<0$.
This can be compared to the behavior of the hysteresis curves determined by~\citet{usov_2012}
when the easy axis of the uniaxial MAE component is fixed relative to the NP frame.
This shoulder is enhanced when either the inverse temperature $\beta$ increases or when the 
polydispersity $\s$ increases (see figure~\ref{shoulder}).
This latter point is simply due to the presence of larger particles 
in the distribution when $\s$ increases, with accordingly larger anisotropy energies. 
We can be interpret this feature as the coherent contributions of uniaxial and cubic terms. 
In the case $\ep_c<0$ where the favorable orientations are the $\{111\}$ axes, 
we find that the cubic contribution remains to enhance the uniaxial anisotropy constant
by a factor of roughly $\abs{\ep_c}/5$ as shown in the inset of figure~(\ref{comb_ek_kc_noddi}).
\subsection* {Interacting systems}
\begin{figure}[h]   
\includegraphics[width = 11.9 cm , angle = -00]{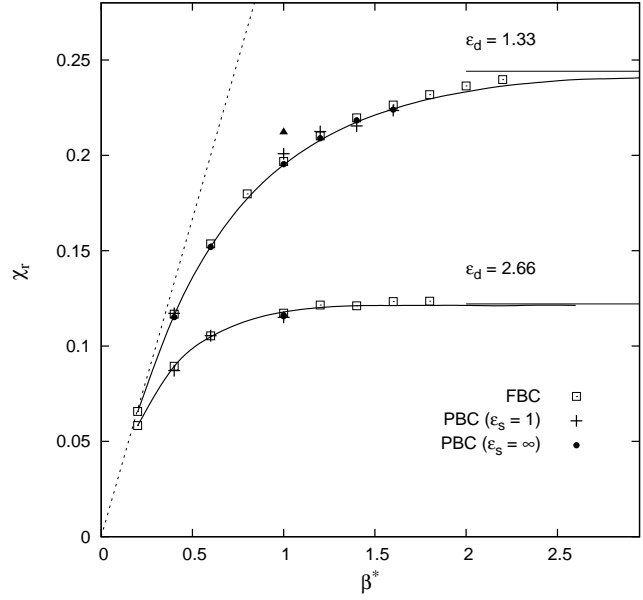}
\caption{\label{fig_chi} 
Reduced linear susceptibility, $\chi_r$ versus the inverse reduced temperature $\be^*$ in 
the quasi modisperse case, $\s~=~0.05$ for a volumic fraction $\phi~=~0.385$, $\ep_d$~=~1.33 and~2.66 
($\ep_d^{eff}$~=~1.0 and 2.0 respectively).
Different boundary conditions are considered. In the PBC with Ewald sums,
the number of particles is $N_p$~=~600 while the clusters for the FBC include $N_p$~=~1000 particles.
Solid line : $M/M_s$ for $h$~=~1. Solid horizontal lines indicate the
limit for $y~\tend~\infty$, (equation~(\ref{chi_lim}).
The solid triangle at $\be^*$~=~1 indicates the value of $\chi_r$ for $\ep_d^{eff}$~=~1.0 in the polydisperse case 
$\s$~=~0.28 ($\ep_d$~=~1.73; $\gD/r_m$~=~0.40).
}
\end{figure}

Most of our simulations with DDI are performed with free boundary conditions (FBC) on large 
spherical NP clusters of $N_p$\,$\sim$\,1000 particles. In order to check the validity of the method, 
we have performed simulations with periodic boundary conditions (PBC) with Ewald sums for the DDI 
in both the conducting or the vacuum external boundary conditions~\citea{allen_1987,frenkel_2002}. 
This is done by using either $\ep_s$=1 or $\ep_s=\infty$ for the surrounding permeability (or 
dielectric constant in the electric dipolar case). Here we are interested in the
determination of the linear susceptibility for the infinite system embedded in vacuum, as we seek 
the magnetic response in terms of the external field. Therefore, we check that one can 
get $\chi_r(\ep_s\,=\,1)$ from simulations on a large spherical NP cluster with FBC,
or by using PBC with Ewald sums in either the conducting or the vacuum boundary conditions. 
The value of $\chi_r(\ep_s\,=\,1)$ can be obtained from a simulation with external conducting 
conditions by exploiting in equation~(\ref{diel_1}) the independence of $\mu$ with respect of 
$\ep_s$ as it is an intrinsic property ,
\equa{chi_inf_vac}
\chi_r(\ep_s = 1) = \chi_r(\ep_s = \infty)/(1 + 8\phi\ep_d\chi_r(\ep_s = \infty)).
\auqe
The comparison of $\chi_r(\ep_s\,=\,1)$ from the three routes is shown in figure~\ref{fig_chi} 
in the absence of anisotropy and in the quasi monodisperse case ($\s$\,=\,0.05).
We have used the same initial cluster and extracted either a spherical cluster for FBC or a 
cubic simulation box for PBC with a value of $\gD$ fitted on the volume fraction $\phi$.
Moreover we have checked that for moderate values of the DDI coupling the permeability obtained 
from these three routes leads to similar values. These two points show the coherence of our 
simulations with DDI. 
When compared to the results of~\citet{klapp_2001} the curve $\mu(y)$ we get at $\phi$\,=\,0.385
lies in between the ones of the frozen model with correlation and of the frozen model with 
quenched disorder, much closer to the former and in fact very close to that of the DHS fluid.

\begin{figure}[h]   
\includegraphics[width = 11.9 cm , angle = -00]{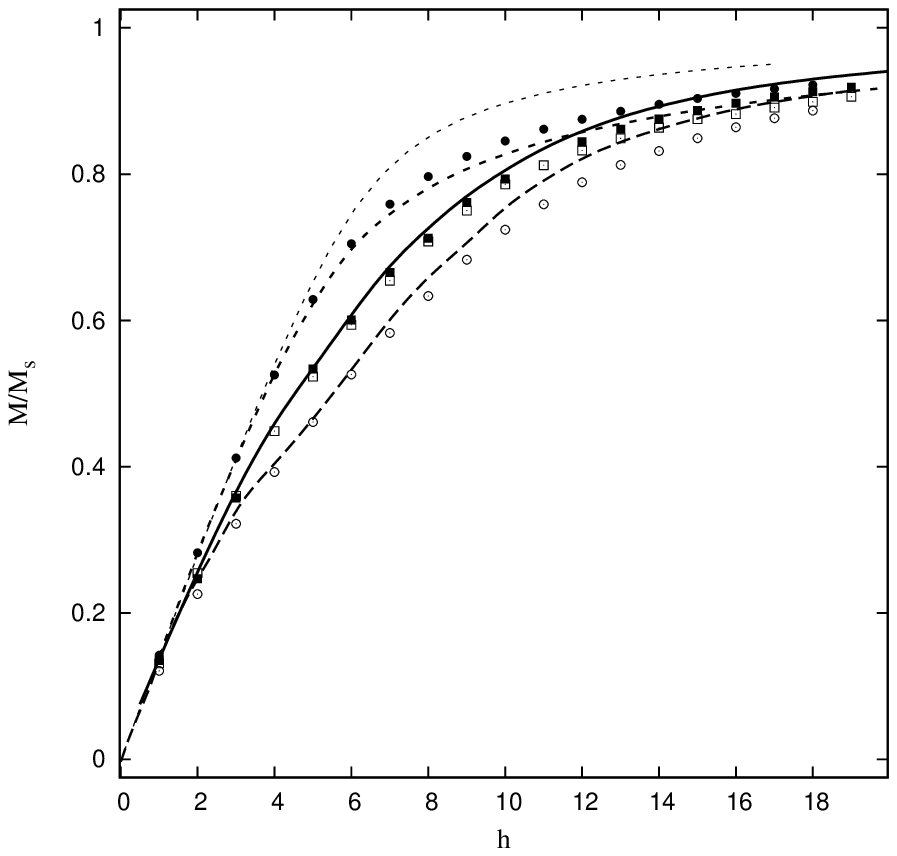}
\caption{\label{sp_ddi_2.37_028}
Reduced magnetization for a polydisperse interacting system with $\be^*$~=~1, $\ep_d$~=~2.37, 
$\gD/d_m$~=~0.15, polydispersity $\s$~=~0.28 and different sets of MAE constants.
$\ep_{uv}$~=~0.0 and $\ep_c$~=~0, dotted line;
$\ep_{uv}$~=~5.0 and $\ep_c$~=~0, solid line;
$\ep_{uv}$~=~5.0, $\ep_c$~=~15  and $\hat{n}$~=~random, open squares;
$\ep_{uv}$~=~5.0, $\ep_c$~=~-15 and $\hat{n}$~=~random, solid squares;
$\ep_{uv}$~=~5.0, $\ep_c$~=~15  and $\hat{n}$~=~[111], short dashed line; 
$\ep_{uv}$~=~5.0, $\ep_c$~=~-15 and $\hat{n}$~=~[111], long dashed line;
$\ep_{uv}$~=~5.0, $\ep_c$~=~-15 and $\hat{n}$~=~[001], solid circles;
$\ep_{uv}$~=~5.0, $\ep_c$~=~15  and $\hat{n}$~=~[001], open circles.
}
\end{figure}

Beside the strong reduction of the initial susceptibility, the DDI reduce also the
deviation of the $M(h)$ curves due to MAE, as can be seen in figure~\ref{sp_ddi_2.37_028}.
As expected the cubic anisotropy has nearly no influence on the $M(h)$ when the
easy axes and the cubic axes are independently randomly distributed; on the other
hand the change in the $M(h)$ curve due to the cubic contribution when
$\{\hat{n}\}_i$ are along the crystallites $[111]$ with $k_c~<~0$ or along the
$[001]$ with $k_c~>~0$ is smaller than in the absence of DDI.
Nevertheless, the contribution of the cubic anisotropy may be not
negligible under the condition of a coherence with the uniaxial term.
Moreover, we do find that in order for the cubic term to give a noticeable effect
a rather large value of the cubic anisotropy constant, $k_c$ is necessary.

\begin{figure}[h]   
\includegraphics[width = 0.75\textwidth , angle = -00]{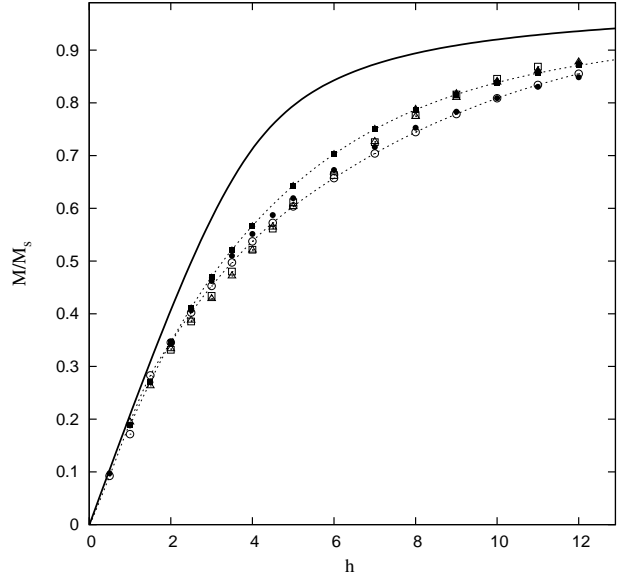}
\caption{\label{ed_eff_1_equiv_an}
Reduced magnetization for the effective DDI coupling constant $\ep_d^{eff}$~=~1.0,
$\be^*$~=~1, polydispersity $\s$~=~0.28, open symbols or $\s$~=~0.05, solid symbols.
$\ep_{uv}$~=~6.30 and $\ep_c$~=~0, circles;  
$\ep_{uv}$~=~4.0, $\ep_c$~=~-12.0 and $\hat{n}$~=~$[111]$, squares;  
$\ep_{uv}$~=~4.0, $\ep_c$~=~50,   and $\hat{n}$~=~$[001]$, triangles.  
$\ep_{uv}$~=~0.0, $\ep_c$~=~0     and $\s$~=~0.28, solid line.
The dotted lines are guides to the eyes.
}
\end{figure}

In opposite to what we get in the absence of DDI, we do not find any distinctive feature 
of either the cubic or the uniaxial symmetry on the $M(h)$ curve if the cubic axes are 
randomly distributed in the case of combined or only uniaxial anisotropy.
This is shown in figure~\ref{ed_eff_1_equiv_an} where different combinations of anisotropies
leading to comparable $M(h)$ curves are considered for $\ep_d^{eff}$~=~1. 

Finally we consider the comparison with the experimental magnetization curves
of Ref.~\citeb{de-montferrand_2012} on powder samples of maghemite NP
differing by their size. These samples are characterized by a polydispersity 
$\s\sim~0.27$ and the estimated coating layer thickness is {\it c.a.} 2 $nm$. 
The behavior of the $M(H_a)$ curve being controlled by the DDI and the MAE at low and intermediate
values of the applied field respectively,
we fit the value of $\gD$ by the slope at $H_a~\sim~0$ and the anisotropy constants on the
behavior of $M(H_a)$ at higher values of $H_a$. 
We find that the region $H_a$~$\sim$~0 is well reproduced with $\gD$~=~2~$nm$ for 
$d_m$~=~10~$nm$ and 21~$nm$, and $\gD$~=~2.4~$nm$ for 12~$nm$, which does not differ much
from the estimated experimental value.
Concerning the cubic anisotropy since the experimental samples are not textured we consider only a
random distribution of cubic axes. 
The value of the corresponding anisotropy constant may differ from its known bulk value due to
surface effects; however, we consider the bulk value as a starting point.
In any case, since the cubic anisotropy constant for iron oxide is rather small, we expect
only a small effect of the cubic contribution to the MAE and accordingly we consider only the
case where the cubic and the uniaxial components of the MAE reinforce each other. With $\ep_c$\,$<$\,0,
this means that we limit ourselves to a easy axes distribution $\{\hat{n}\}_i~=~[111]_i$.
For the uniaxial MAE we have to choose either a surface or volume dependent MAE
(see equation~(\ref{ener_dip2})); however, we have shown that the surface dependent MAE can be 
reproduced by the volume dependent one through the effective constant of~(\ref{kv_eff}). Hence, 
starting from the bulk value for $\ep_c$ we are left with $\ep_{uv}$ as the only fitting parameter.
We find $\ep_{uv}$~=~4.00 for $d_m$~=~10\,$nm$ by fitting $M(H_a)$ in the intermediate field range;
then, the same quality of agreement between the model and the experimental curves is obtained
for $d_m$~=~12\,$nm$ and 21\,$nm$ by using a value of $\ep_{uv}$ scaling as $d_m^3$, namely
$\ep_{uv}$~=~6.912 and 32.0 for $d_m/d_{ref}$~=~1.2 and 2.0 respectively,
{\it i.e.} $K_v$~=~31.6~kJm$^{-3}$ (we use the simulated curve for
$d_m/d_{ref}$~=~2 for comparison of the experimental curves of the samples with $d_m$~18 and 21~$nm$; 
only the second is presented here). Notice the weak hysteresis cycle for the experimental sample 
characterized by $d_m$~=~21nm; this is due to the largest particles in the distribution and is not
reproduced by the M.C. simulations, since we have chosen to perform equilibrium ($\tau_m$~=~$\infty$)  
simulations only.
The cubic MAE gives only a small contribution to $M(H_a)$ as 
illustrated by the difference obtained using $\ep_c$ deduced from either the magnetite or the 
maghemite bulk values given in Table~\ref{tab_para} 
(see figures~\ref{comp_exp12nm} and \ref{comp_exp21nm}).
Therefore, we find that using the iron oxide bulk value for the cubic MAE constant
the experimental NP of Ref.~\citeb{de-montferrand_2012} can be modeled excepted in the high field 
region, by NP presenting a volume dependent uniaxial anisotropy with $K_v$~=~31.6\,kJm$^{-3}$.
However, as we have shown, we can get similar $M(H_a)$ curves with different combinations of
cubic and uniaxial MAE especially with the DDI which weaken the peculiar features of the cubic 
contribution. Hence, we can get the same agreement with experiment by using on the one hand a
uniaxial MAE scaling as $d_m^2$ corresponding to a surface anisotropy and on the other hand
a fitted cubic contribution. Starting from $\ep_{uv}$~=~4 for $d_m/d_{ref}$~=~1.0 this gives
$\ep_{uv}$~=~5.76 for $dm/d_{ref}$~1.2 (which translates to $\ep_{us}$~=~7.03 for $\s$~=~0.28
and $K_s$~=~6.45\,10$^{-5}$~Jm$^{-2}$). The corresponding cubic component is obtained from our 
finding that an increase of $\abs{\ep_c}$ corresponds to an increase of $\ep_{uv}$ of 
roughly $\abs{\ep_c}/5$, leading to $\ep_c$~=~-9 and $K_c$~=~-41\,kJm$^{-3}$. 
We have also considered a fitted cubic MAE with a positive $\ep_c$, and $\hat{n}_i$~=~$[001]_i$
for which we find $\ep_c$~=~5.0($K_c$~=~25.15~kJm$^{-3}$).
The results is shown in figure~\ref{comp_exp12nm}. Doing this means that the cubic 
anisotropy energy present an anomalous component, namely $\abs{K_c-K_c^{bulk}}$, 
scaling as the NP volume while it should be understood as a surface effect.
Hence, although it seems difficult to conclude on the best fit of the experimental set considered,
it may be better to avoid the latter contradiction and consider these NP as presenting a volume
dependent uniaxial MAE; however, we then get a value for the effective anisotropy constant
too large to be explained only as a shape anisotropy. It is nevertheless still in the range
of what is obtained experimentally from $T_B$ for iron oxide NP.
In any case, we have to take
such conclusions with care given the simplicity of the model. Similarly, the high field range cannot
be reproduced with the simple OSP model and necessitates a the inclusion of a field dependent
description of the individual NP.

\section {Conclusion}
\label {conclusion}
In this work, we have performed Monte Carlo simulations of room temperature magnetization
curves in the superparamagnetic regime, with a particular attention paid to the iron oxide 
based NP. We focused on the search for a peculiar feature of the cubic MAE component on 
the $M(H_a)$ curve since iron oxide and spinel ferrites in general presents an intrinsic
MAE with cubic symmetry while from experiments a uniaxial MAE is generally found.
Our result is that a peculiar feature of the cubic component can be obtained only
{\it i)}~if the the cubic and the uniaxial components are correlated through the alignment 
of the NP easy axes on a specified crystallographic orientation of the crystallites;
{\it ii)}~if the DDI are negligible {\it via} a small NP volume fraction.
Nevertheless a large value of the cubic MAE constant compared the uniaxial one is necessary
for the former to give a noticeable effect on the room temperature $M(H_a)$.
\renewcommand {\theequation }{\Alph{comptapp}.\arabic{equation}}

\setcounter {equation}{0} \setcounter {comptapp}{\numexpr\value{comptapp}+1}
\section {Appendix~\thecomptapp}
\label {app_phi}

\noindent
In this appendix we explicit the function $\varphi$ introduced in 
equation (\ref{ed_eff}). The volume fraction is defined as 
\equa{phi_1}
\phi = N_p \frac{1}{V} \int_0^{\infty} f(d) \frac{\pi}{6} d^3 d(d)
    = \frac{\pi d_m^3}{6 V} d_3^*
\auqe
where $V$ is the total volume and $d_n^*$ is the reduced $n-th$ moment of $f(d)$.
Each particle of diameter $d$ is surrounded by a coating layer of thickness $\gD/2$; the
maximum value of the volume fraction, $\phi_m$ is obtained as the volume fraction of
the spheres including both the particles and the coating layer, namely by replacing
$d$ in (\ref{phi_1}) by $(d\;+\;\gD)$ with the same distribution function.
Defining $\gD^*\;=\; \gD/d_m$ we get 
\equa{phi_2}
  \phi_m &=& N_p \frac{1}{V} \int_0^{\infty} f(d) \frac{\pi}{6} (d + \gD)^3 d(d) \nonumber \\
         &=& \frac{\pi d_m^3}{6 V} d_3^* (1 + \gD^*)^3 
  \left [ \frac{ 1 + 3 \gD^* (d_2^*/d_3^*) +3 \gD^{*2} (d_1^*/d_3^*) + \gD^{*3} (1/d_3^*) } 
               {\left( 1 + \gD^* \right)^3} \right]
\auqe
which defines the function $\varphi$ as the expression in square brackets. From the analytical
expression of the reduced moments $d_n^*$ in the lognormal law, 
$d_n^*\;=\;\exp(n^2\s^2/2)$ we get
\equa{phi_3}
\varphi = \left [ \frac{1 + 3\gD^{*} e^{-5\s^2/2} + 3\gD^{*2}e^{-4\s^2} + \gD^{*3}e^{-9\s^2/2} } 
          {\left( 1 + \gD^* \right)^3} \right]
\auqe

\gdsaut
\section* {Acknowledgments}
This work was granted access to the HPC resources of CINES under the allocation 
2013-c096180 made by GENCI (Grand Equipement National de Calcul Intensif).
%

\newpage
%
\begin{figure}[h]   
\includegraphics[width = 0.75\textwidth , angle = -00]{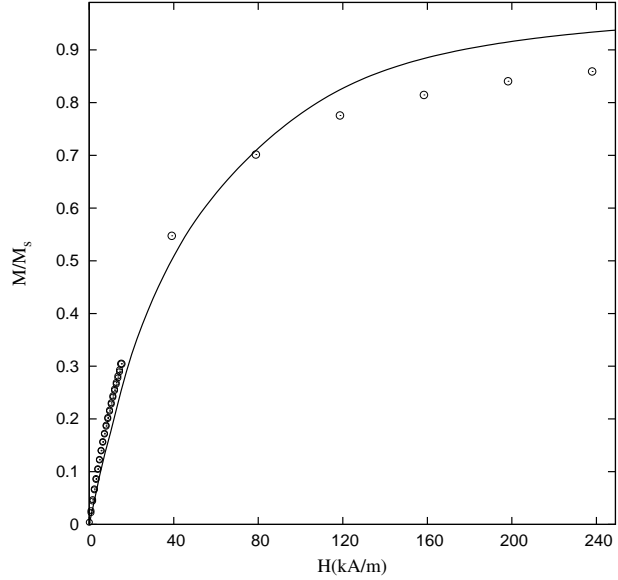}
\caption{\label{comp_exp10nm}
Comparison of the experimental reduced magnetization curve of a maghemite powder
sample~\citea{de-montferrand_2012} with $d_m$~=~10~$nm$, open circles
with the M.C. simulation, solid line. The parameters used in the MC simulation are 
$\s$ = 0.28, $\ep_d$~=~1.0, $\gD/d_m$~=~0.20, $\ep_{uv}~=~4.0$ and $\ep_c~=~-1.5$ with 
$\hat{n}_i~=~[111]$. $\be^*$~=~1.
}
\end{figure}
\begin{figure}[h]   
\includegraphics[width = 0.75\textwidth , angle = -00]{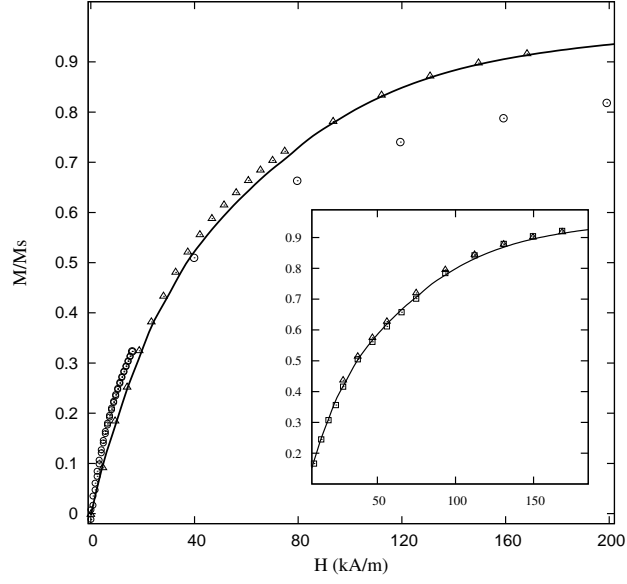}
\caption{\label{comp_exp12nm}
Same as figure \ref{comp_exp10nm} for $d_m$~=~12~$nm$. 
Experiments~\citea{de-montferrand_2012}, open circles. 
The M.C. simulations are performed with
$\s$~=~0.28, $\ep_d$~=~1.733, $\gD/d_m$~=~0.20 and different sets of MAE parameters.
$\ep_{uv}~=~6.912$, $\ep_c~=~-2.85$ and $\hat{n}_i~=~[111]$, solid line.
$\ep_{uv}~=~6.912$, $\ep_c~=~-1.1$ and $\hat{n}_i~=~[111]$, open triangles.
Inset: Comparison of the simulated $M(H_a)/M_s$ curves with
$\ep_{uv}~=~6.912$, $\hat{n}_i~=~[111]$ and $\ep_c~=~-2.85$, solid line;
$\ep_{uv}~=~5.76$, $\hat{n}_i~=~[111]$ and $\ep_c~=~-9.00$, open triangles;
$\ep_{uv}~=~5.76$, $\hat{n}_i~=~[001]$ and $\ep_c~=~5.50$, open squares.
}
\end{figure}
\begin{figure}[h]   
\includegraphics[width = 0.75\textwidth , angle = -00]{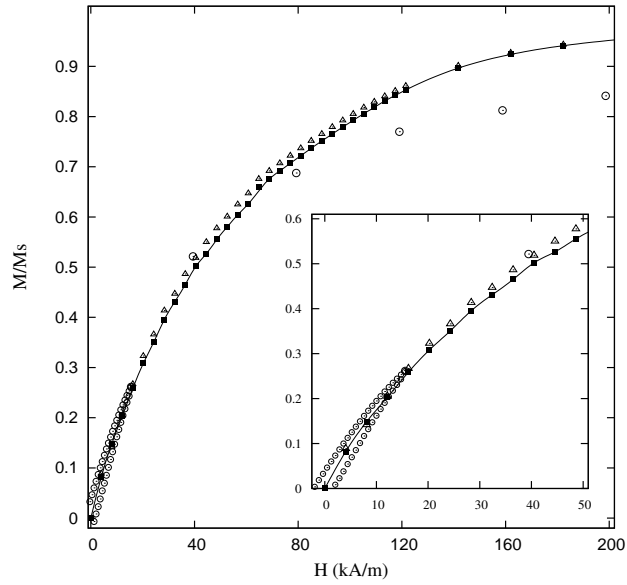}
\caption{\label{comp_exp21nm}
Same as figure \ref{comp_exp10nm} for $d_m$~=~21~$nm$.
Experiments~\citea{de-montferrand_2012}, open circles. 
The M.C. simulations are performed with $d_m/d_{ref}$~=~2,
$\s$~=~0.28, $\ep_d$~=~8.0, $\gD/d_m$~=~0.10, $\ep_{uv}~=~32.00$, 
$\hat{n}_i~=~[111]$ and $\ep_c$~=~-13.2, solid squares or $\ep_c$~=~-5.0, open triangles. 
The thin solid line is a guide to the eyes. 
}
\end{figure}
%
\end {document}